# A Tutorial of Cyber-Syndrome viewed from Cyber-Physical-Social-Thinking Space and Maslow's Hierarchy of Needs

Feifei Shi[1], Huansheng Ning[1], *Senior Member*, IEEE, Sahraoui Dhelim[2]

**Abstract**

With the increase of active Internet users, various physical, social, and mental disorders have recently emerged because of the excessive use of technology. Cyber-Syndrome is known as the condition that appears due to the excessive interaction with the cyberspace, and it affects the users' physical, social, and mental states. In this paper, we discuss the etiology and symptoms of Cyber-Syndrome according to theories of Cyber-Physical-Social-Thinking (CPST) space and Maslow's Hierarchy of Needs. In addition, we also propose an entropy-based mechanism for recovery of Cyber-Syndrome, to provide potential guidance for clinical detection and diagnosis. Cyber-Syndrome has attracted much attention these days, and more in-depth exploration is needed in the future.

## I. Introduction

The wide spread of cyberspace and cyber techniques has brought significant benefits to our daily life and industrial manufacturing [1]. As reported by Statista, up to January 2021 there were almost 4.66 billion active Internet users all over the world [2]. Most people are enjoying convenience of the Internet, taking various activities such as playing games, watching videos, and online shopping. In recent years, due to the travel restrictions related to Covid-19 pandemic, many international cooperation and businesses have been gradually shifting to online forms. The Internet is no longer just a tool for entertainment, but a necessity for work and daily routines [3].

---

[1] School of Computer and Communication Engineering, University of Science and Technology Beijing, 100083, Beijing, China.

[2] School of Computer Science, University College Dublin, Belfield, Dublin 4, Ireland.

However, the overwhelming penetration of the Internet also leads to a series of problems that cannot be ignored, such as Internet addiction [4] and problematic Internet use [5], which will affect physiological and psychological health to a large extent. In 2016, we once made some investigations between Internet use and physical complaints [6]. In 2018, we initially concluded the emerging cyber-enabled disease as Cyber-Syndrome, with physical, social, and mental disorders [7]. It is the first to regard Cyber-Syndrome as a whole and has provided a relatively general description of Cyber-Syndrome ranging from definition, classification, recovery, and prevention. The work opens a new way of cyber health and disease and lays a solid foundation for further research. Based on previous work, we have continued much in-depth research in the following years and generated some new ideas around Cyber-Syndrome.

As we all know, the Internet and cyber techniques enable almost everything in physical, social, and thinking space to be connected. It has largely hastened the paradigm of Internet of X and an unprecedented convergence of Cyber-Physical-Social- Thinking (CPST) space [8], [9]. Although Cyber-Syndrome is an emerging new disease of cyberspace, similar like Parkinson's disease in physical space, it not only stems from factors in cyberspace, but is also indirectly affected by physical, social, and thinking spaces.

At the same time, we notice the theory of Maslow's Hierarchy of Needs. It reveals the motivations of humans' behaviors from the psychological perspective [10]. In 1940s, Maslow first established a five-tier model composed of physiological needs, safety needs, love and belonging needs, esteem needs, and self-actualization needs. In the following twenty years, the hierarchical theory has been expanded including cognitive needs, aesthetic needs and transcendence needs [11]. Maslow's Hierarchy of Needs analyzes psychological motivations for specified behaviors to al certain extent, for example, working hard to gain a sense of psychological security. It inspires us to expound the complicated etiology of Cyber-Syndrome from a more intuitive perspective of psychological motivations.

In this paper, we have made some extensions and improvements based on previous work and present a tutorial of Cyber-Syndrome with theories of CPST space and Maslow's Hierarchy of Needs. To prompt the dissemination, we first analyze its etiology based on the theories of CPST space and Maslow's Hierarchy of Needs. Following that, the cyber-enabled symptoms of Cyber-Syndrome are illustrated according to physical, social, and thinking spaces respectively. Moreover, we propose an entropy-based mechanism aimed at the recovery of Cyber-Syndrome. The main highlights of this paper include:

- Illustrating the etiology of Cyber-Syndrome with theories of CPST space and Maslow's Hierarchy of Needs to strengthen the comprehension of Cyber-Syndrome.

- Concluding and analyzing the symptoms of Cyber-Syndrome from cyber-enabled physical, social and thinking spaces, thus, to provide potential guidance for early detection in practical applications.

- Proposing an entropy-based mechanism of Cyber-Syndrome recovery based on CPST space and making available references in clinical diagnosis.

The remainder of this paper is arranged as follows. Section II lists related concepts involved in this paper and presents a comparison between similar concepts to Cyber-Syndrome. Section III analyzes the influential factors of Cyber-Syndrome on the basis of CPST space and Maslow's Hierarchy of Needs, and comprehensively introduces its etiology. Section IV depicts the cyber-enabled symptoms of Cyber-Syndrome in CPST space respectively. Section V proposes an entropy-based mechanism for Cyber-Syndrome recovery. Section VI gives a conclusion and outlines further directions of Cyber-Syndrome.

## II. RELATED CONCEPTS

As Cyber-Syndrome is a complicated disease that mainly stems from the unhealthy Internet use, we emphasize its etiology, symptoms, and recovery with regarding to the CPST space and Maslow's Hierarchy of Needs. In this section, we give an explicit introduction around the related concepts and works.

### A. Cyber-Syndrome

In 2018, we discussed the concept of Cyber-Syndrome, and illustrated its definition with the etymology of "Cyber" and "Syndrome" [7]. Syndrome refers to sets of medical symptoms and signs that are along with one or more diseases or disorders. The work first concludes Cyber-Syndrome from its definition, classification, recovery, and prevention, and has inspired many follow-up studies. Based on previous work, we conclude similar works to Cyber-Syndrome, and makes a comparison around definition and symptoms as shown in Table I.

**Table I. Some Typical Concepts Around Health in Cyberspace.**

| Name | Definition | Symptoms | References |
|---|---|---|---|
| Internet addiction disorder | It describes the disorder due to excessive Internet use and is accompanied by withdrawal effects after cessation of Internet use. | Psychological dependence, compulsive use, emotional changes, and avoidance of work etc. | [12]-[14] |
| Cyber disorders | The mental health concern associated with significant social, psychological, and occupational impairment. | Loss of self-esteem, self-confidence, and sense of security; Types of cyber sexual addiction, cyber bullying, cyber-relationship addiction, | [15], [16] |

| | | computer addiction, and Internet addiction etc. | |
|---|---|---|---|
| Cybersickness | It represents the negative feelings users suffer from during or after immersion into virtual reality. | Nausea, headache, dizziness, and sweating etc. | [17]-[19] |
| Cyberchondria | It represents the unfounded escalation of health concerns during web search. | Overwhelming anxiety, unexpected investment of time, as well as expensive engagements with healthcare professionals. | [20], [21] |
| Cyber-Syndrome | It is the physical, social, and mental disorders due to excessive cyberspace interaction. | Tactile sensation, radiation exposure, behavior disorders, mood disorders, social anxiety, social hostility etc. | [7] |

*1) Internet addiction disorder:* Internet addiction disorder (IAD), the most familiar concept around cyber health, emphasizes the psychological dependence on Internet or electronic devices. It mainly stems from the problematic or compulsive Internet use and will suffer from withdrawal symptoms after caseation [12]-[14]. Although there may be physical and social complications around Internet addiction, the most obvious symptom is the psychological addiction to Internet, with losing the ability to control Internet time use.

*2) Cyber disorders:* A group of Cyber disorders concerns more about the influence on mental health, with social, psychological, and occupational impairment [15]. In 2015, Palanichamy introduced five types of cyber disorders, namely cyber bullying, cyber sexual addiction, cyber-relationship addiction, computer addiction, and Internet addiction [16]. Compared with IAD, the concept of Cyber disorders is much broader, and includes the symptoms and signs brought by IAD.

*3) Cybersickness:* Cybersickness is a kind of motion sickness that results from the exposure to cyberspace, especially in virtual environments [17], [18]. As with motion sickness, humans who are with cybersickness are sometimes stimulated by cyber techniques or equipment, and confused with symptoms like nausea, headache, dizziness, and sweating etc. For example, if users scroll down the screen quickly, they may feel dizzy to some extent. The most prevalent remedy is to stop long-time cyber exposure and have a rest if the cybersickness strikes.

*4) Cyberchondria:* Cyberchondria refers to the unfounded escalation of healthy concerns when searching for information online [20], [21]. Different from the concepts mentioned above, Cyberchondria stands for the influence brought by online information overload, particularly when searching for relative information around health. Patients with Cyberchondria may experience slight or heavy anxiety with reference to complicated information and hints in cyberspace.

Compared with the concepts discussed in this section, Cyber-Syndrome is a much more complicated disease in cyberspace. It results from the excessive exposure and interactions in cyberspace, meanwhile has explicit symptoms in cyber-enabled physical, social, and thinking space. Last years has witnessed the rapid development of Cyber-Syndrome, and it would ruin humans' daily life to a great extent with no efficient intervention.

*B. Cyber-Physical-Social-Thinking space*

The advances of Internet and cyber techniques make things in physical, social, and thinking spaces to be connected. In 2015, Ning established a novel Cyber-Physical-Social-Thinking (CPST) space, and provides a general blueprint where humans live [8]. With the prevalence of cyberspace and cyber techniques, CPST space has been gradually recognized as basic living space for humans.

As can be seen in Figure 1, the CPST space is depicted in a direct way. Physical space, the traditional space where we are living, is composed of humans, animals, plants, buildings, atmosphere, and other environmental elements. It could be regarded as the primary living space for humans, and is able to satisfy their needs of water, food, warmth and so forth.

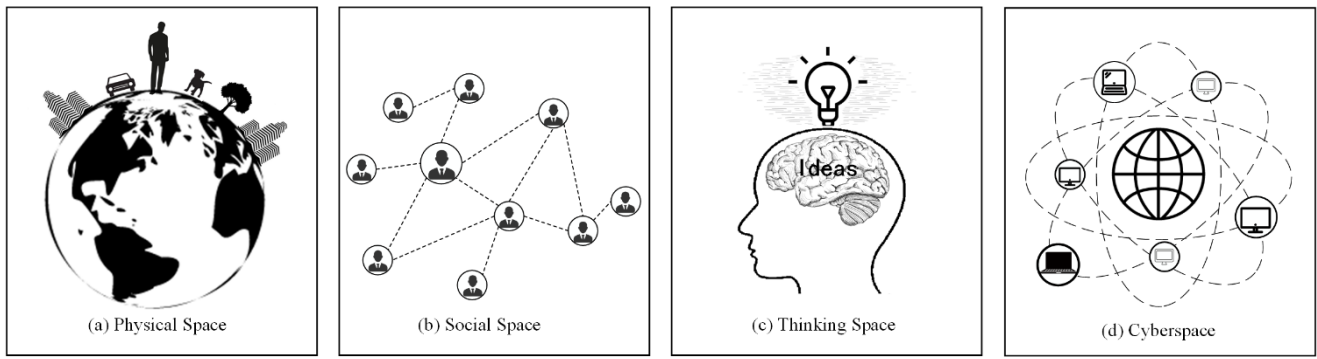

**Figure 1. The Cyber-Physical-Social-Thinking space [8].**

Social space, as name implies, is established with relationships among humans. The relationships may be built upon geographical locations, blood, and social reasons etc. These years with the development of the Internet, humans are tightly connected with various electronic devices, and are playing much more central roles under a new paradigm of Internet of People (IoP) [22].

Thinking space is the mental world in marvelous brains. As we all know, humans' brains are superior to those of animals due to complicated mechanisms and tissues. They have independent thinking spaces full of wonderful ideas, thoughts, and creative abilities. It also makes humans could think intelligently when dealing with complex problems.

Cyberspace, obviously, is the virtual world brought by Internet. It depicts a cyber world where humans could carry on various activities such as entertainment, study and work online. It is reported that an increasing number of users spend more time on Internet activities, while it also causes series of challenges in physical, social, and thinking spaces. For example, the long-term Internet use may give rise to physical discomforts, as well as the reduction in social time face to face.

In a word, the CPST space draws a blueprint of basic living space for humans. Inspired by it, we regard Cyber-Syndrome is not a simple disease only in cyberspace, and should be a complicated syndrome influenced by physical, social, thinking, and cyber spaces. Therefore, we are going to analyze Cyber-Syndrome with reference to CPST space, including the etiology, symptoms as well as the recovery mechanism.

*C. Maslow's Hierarchy of Needs*

In 1943, the theory of Maslow's Hierarchy of Needs was first proposed with a five-tier pyramid, including physiological needs, safety needs, love and belonging needs, esteem needs, and self-actualization [10]. After twenty years, with the progress in social psychology, the hierarchy of needs has been expanded with cognitive needs, aesthetic needs, and transcendence needs [11]. The theories of Maslow's Hierarchy of Needs describe motivations of humans and are used to explain daily actions or behaviors. In this paper, we adopt the five-tier Maslow's Hierarchy of Needs to help understand Cyber-Syndrome.

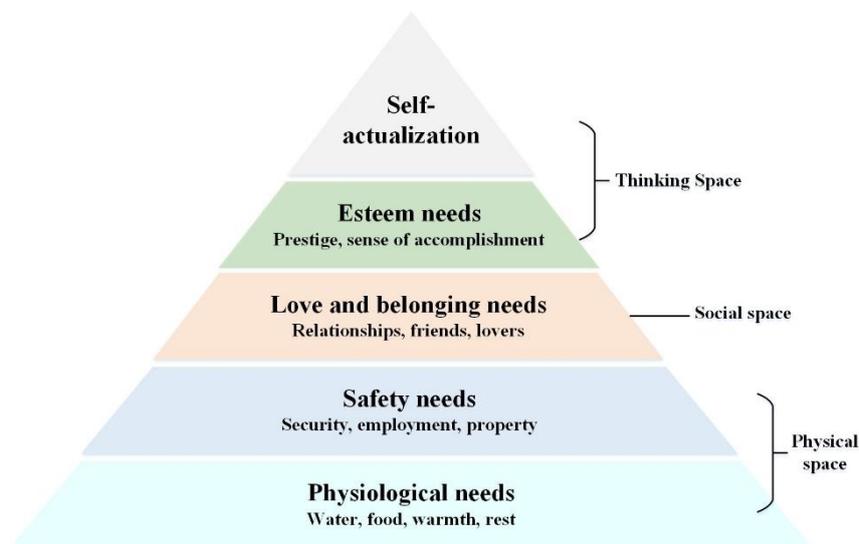

Figure 2. The Five-tier Maslow's Hierarchy of Needs [10].

As can be seen in Figure 2, the five-tier Maslow's Hierarchy of Needs has been depicted from basic level of needs to higher-level needs. The basic level of needs includes physiological needs and safety needs such as water, food, warmth, private safety and so forth. These are primary requirements needed to be satisfied for humans. Apart from those basic needs, the motivations like love and belonging, esteem, and self-actualization are more advanced. Humans would like to pursue these higher-level needs in daily life to realize self-affirmation and self-worth.

In Figure 2, we also try to connect the theory of Maslow's Hierarchy of Needs with Physical-Social-Thinking (PST) space. The basic physiological needs and safety needs could be marked as motivations in physical space, to guarantee a healthy and safe living environment. The love and belonging needs belong to social space, as humans wish to build friendly and close relationships with others, to achieve love and sense of belonging. The upper levels of esteem and self-actualization needs are motivations of self-fulfillment, and they are correspondent to requirements in thinking space. People who have strong desires of personal development will strive better to satisfy the advanced needs.

As Maslow pointed out the man is always a wanting animal, the hidden motivations to satisfy specific requirements or needs could translate humans' behaviors to some extent. Hence it is possible to explain and understand Cyber-Syndrome based on Maslow's Hierarchy of Needs.

### III. THE ETIOLOGY OF CYBER-SYNDROME FROM CPST SPACE AND MASLOW'S HIERARCHY OF NEEDS

Obviously, Cyber-Syndrome, the sets of signs and symptoms, are directly influenced by excessive or problematic Internet use. However, the etiology of Cyber-Syndrome is so complicated, and cannot be easily explained as a simple disease in cyberspace. In order to provide a more understandable analysis of Cyber-Syndrome, we discuss its etiology combined with CPST space and Maslow's Hierarchy of Needs.

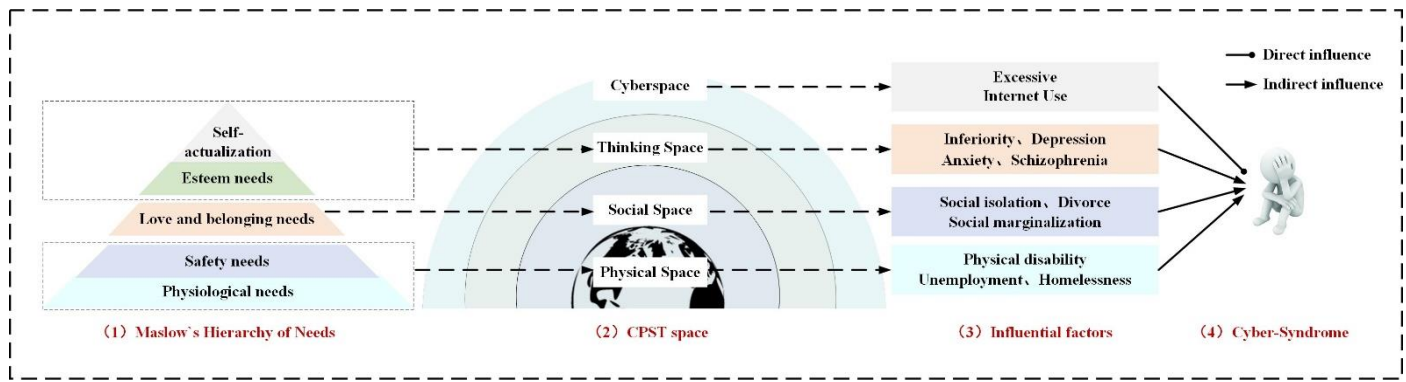

Figure 3. The Etiology of Cyber-Syndrome based on CPST space and Maslow's Hierarchy of Needs.

In Figure 3, we present the etiology of Cyber-Syndrome around living space of CPST space. As we discussed before, Cyber-Syndrome not only results from unhealthy interactions with cyberspace, but also affected by factors in physical, social, and thinking spaces. Based on the theory of the five-tier Maslow's Hierarchy of Needs, we clearly interpret the etiology of Cyber-Syndrome in each space by analyzing the influential factors.

In physical space, humans need to satisfy basic requirements, such as physiological needs of water, food, and warmth for primary living, safety needs of property and employment etc. It encourages humans to work hard to guarantee primary living environment. However, people may suffer from physical disability, unemployment, homelessness, and other painful experience, and fail to meet the expectations in physical space. The frustration in real life may lead people to turn their attention to cyberspace, resulting in excessive Internet use.

In social space, humans desire to attain love and sense of belonging. On the basis of physiological and safety needs, people tend to establish relationships with others, and acquire the sense of belonging, including affiliation, trust, friendship, intimacy, and affection etc. Nevertheless, social isolation, deception, divorce, and social marginalization may be obstacles in achieving love and belonging. Therefore, people who are dissatisfied with their social relationships or social environments are more likely to have evasive emotions and throw themselves to Internet and cyberspace.

As for thinking space, humans are superior in thoughts, ideas, and creativity. They are born to expect attention, recognition, and prestige, to achieve self-actualization. However, people in daily life may be confronted with various challenges and thus lose confidence in themselves. Sometimes they may be troubled by mental illnesses such as depression, anxiety, and schizophrenia etc. At this stage, they may feel more comfortable when addicting into cyberspace, and would like to spend more time in the virtual world.

To sum up, Cyber-Syndrome is a complicated disease jointly impacted by factors in physical, social, thinking, and cyber spaces. Usually, humans who fail to satisfy Maslow's Hierarchy of Needs in real life, are more likely to indulge in cyberspace, and easy to get sick with Cyber-Syndrome. They may have obstacles in maintaining a healthy and safe living environment or establishing and keeping comfortable relationships with others. All those challenges would make humans lose confidence in daily life, and in turn aggravate symptoms of Cyber-Syndrome.

## IV. THE CYBER-ENABLED SYMPTOMS OF CYBER-SYNDROME

In order to better understand Cyber-Syndrome, it is also important to comprehensively recognize its symptoms. In our work of 2018, we have already mentioned the classification of Cyber-Syndrome including physical, social, and mental disorders, as well as different complications. Based on the preliminary work, we further discuss and conclude the cyber-enabled symptoms of Cyber-Syndrome in CPST space respectively. Figure 4 presents the symptoms of Cyber-Syndrome in different spaces.

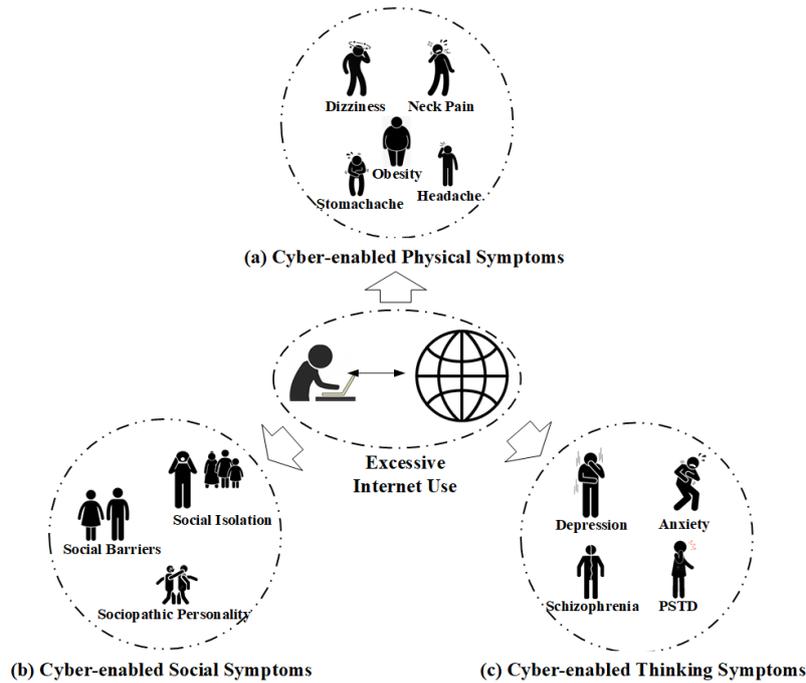

Figure 4. The Cyber-enabled Symptoms of Cyber-Syndrome.

First of all, Cyber-Syndrome has the most explicit symptoms of unhealthy interactions with cyberspace or cyber techniques. Sometimes it may develop towards an overwhelming addiction when humans suffer from severe Cyber-Syndrome. At this stage, people may spend more time in cyberspace and are vulnerable to a series of risks such as cyber bullying, cybercrime and cyber fraud. Cyber bullying depicts the detrimental effects by posting or sending personal information over digital space. It may lead victims to embarrassment and humiliation. Recently there is some research concerning about early detection of cyber bullying, for example, Kumari establishes a model based on VGG 16 network and Convolutional Neural Network to extract both features from images and texts, which improves the performance of cyber bullying early detection [23]. Additionally, cyber fraud also takes place over cyberspace. It is committed to corrupt individual's personal information for benefits, such as identity theft, shipping fraud and so forth. Owolafe once deployed a Long Short-Term Memory-Recurrent Neural Network to detect credit card fraud, and it was demonstrated with a accuracy of 99.6% in financial transaction [24].

In addition to cyberspace, we regard that the Cyber-Syndrome is also reflected in physical, social, and thinking spaces with different features. In physical space, humans who suffer from Cyber-Syndrome may show up physiological symptoms such as dizziness, neck pain, headache, and obesity etc. Since those people would like to keep the same body posture for a long time, it may impact the healthy states of head, neck, finger, back, eyes and so forth. For instance, it has been proved that the prolonged screen time has direct impacts on myopia particularly among infants, children, and adolescents [25]. In addition, the problematic Internet use also triggers problems related to the duration and quality of sleep, as well as the cardiovascular systems [26]. In general, the excessive use of the Internet will take up most of the daily time and will inevitably have an impact on humans' physical conditions.

In social space, Cyber-Syndrome would also present quite some cyber-enabled social symptoms. For example, most people with Cyber-Syndrome fail to satisfy the love and belonging need and may be confused with social relationships and environments. They may escape from the dissatisfied environments and throw themselves into the virtual space, which in turn worsen the situations in real life. By establishing virtual relationships instead of face-to-face communications, the kind of loneliness may increase to some extent, and prevent people to participate in various social activities. In 1998, Tamaki Saito published a book named Social Withdrawal [27]. It defined the concept of social withdrawal which usually takes place when escaping from social activities for more than six months. Following it, Tateno studied and found that there was an overwhelming influence between Internet addiction and social withdrawal [28]. Some other symptoms such as social barriers, and sociopathic personality would also appear, and a vicious circulation would be formed between Cyber-Syndrome and the social symptoms.

Last but not least, Cyber-Syndrome also generates symptoms in thinking space. When it comes to thinking space, the most interesting elements include emotions, mental states, and the thinking abilities. The kind of emotions such as loneliness, depression, and anxiety would usually become obvious with Cyber-Syndrome. For example, Shekar points out that there is a strong relationship between the overwhelming online information and anxiety, and the easily available online information played a catalytic role on anxiety particularly during Covid-19 [29]. Pu provides a systemic research around healthy rumors in Covid-19, and those who are easily to be "infected" by rumors would feel anxious and panic, and some even lose the ability to think independently [30]. More seriously, Todd found that the addiction to Internet was such a craving behavior that may suffer from changes in brain frontal lobe structural stated [31], which would impact the ability of humans to filter information and make reasonable decisions [32].

As mentioned above, we analyze the symptoms of Cyber-Syndrome according to the theories of CPST space. Although Cyber-Syndrome mainly results from the excessive Internet use in cyberspace, it also shows features in physical, social, and thinking spaces. If someone who is experiencing the long-term cyberspace interaction and has presented some similar symptoms discussed in this section, he or she may be confused with Cyber-Syndrome.

## V. AN ENTROPY-BASED MECHANISM OF CYBER-SYNDROME RECOVERY

Cyber-Syndrome is still being regarded as a new emerging disease with less comprehensive knowledge to its early detection and diagnosis, while it is quite important to learn about its etiology, symptoms, as well as recovery methods. In this section, we propose an entropy-based mechanism for Cyber-Syndrome recovery, which may provide guidance in real life.

According to the second law of thermodynamic, the term entropy describes the state of randomness and disorders. It is said that an enclosed system would tend to become more and more chaotic with higher entropic states if there is no external input. Aoki initially proposes theories of entropy in human body, which could also be regarded as a thermodynamic system [33]-[35]. In other words, humans are prone to develop from healthy states to sub-healthy or unhealthy states if there is no intervention, with entropy developing towards higher states.

Therefore, people who get sick with Cyber-Syndrome are also with higher entropic states. They may present symptoms in different CPST space as we discussed in section IV such as tiredness, fatigue and pains in physical space, depression, and anxiety in thinking space. They are easier to catch a cold with weak immune systems and may suffer from series of complications in daily life.

To recover from Cyber-Syndrome, people need to take actions to reduce the entropy production in human body to keep a relatively healthy state. We would like to discuss an entropy-based recovery mechanism of Cyber-Syndrome from the view of CPST space.

As Prigogine mentioned in his Nobel lecture, an entropy formulation was proposed for the irreversible processes in the second law of thermodynamic [36]. The total production of entropy could be calculated with that exchanged with the inside and the outside:

$$dS = d_i S + d_e S \tag{1}$$

Here, the $d_i S$ refers to the internal entropy change while the $d_e S$ represents the external entropy change. For humans, the internal entropy change mainly represents that takes place inside the body, and the external one refers to that takes place between the human body and the respective CPST space. Hence the total

entropy production of human body could be expressed as following:

$$dS = d_iS + \sum_{x \in \{c, p, s, t\}} d_eS_x \tag{2}$$

According to the second law of thermodynamic, the $d_iS$ remains positive since any thermodynamic system would develop towards higher entropic states.

$$d_iS \geq 0 \tag{3}$$

As can be seen in Figure 5, we depict the entropy exchange of the human body with the inside and the outside. In order to express it more conveniently, we use $p(x,t)$, $s(x,t)$, $t(x,t)$, and $c(x,t)$ as the entropy change in physical, social, thinking and cyber spaces, to replace $d_eS_p$, $d_eS_s$, $d_eS_t$, and $d_eS_c$ respectively in following description.

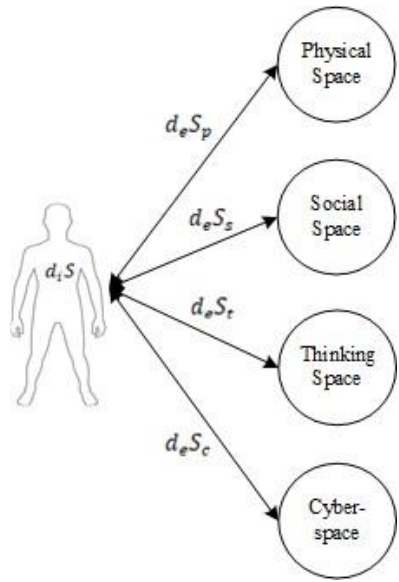

**Figure 5. The Entropy exchange of human body between the inside and the outside.**

To the best of our knowledge, people with Cyber-Syndrome may present unhealthy states from the body inside, hence the value of $d_iS$ keeps going up. In order to keep the balance of healthy states, people with

Cyber-Syndrome need to take actions to slow down the increase of total entropy and try to keep healthy.

However, there are few actions or activities that have an independent influence on a single space. For example, people who supplement a lot of protein will reduce the entropy in the physical space. At the same time, it will also produce energy to do meditation in thinking space or taking part in social activities in social space. Thus, it is not an independent action or activity, and most would impact the entropy in all CPST space. We define any action of humans as $x_i$. For any action of $x_i$ where $i \in \{1,2,3,...,n\}$, there would be corresponding entropy influences of $p_i(x_i, t)$, $s_i(x_i, t)$, $t_i(x_i, t)$, and $c_i(x_i, t)$ in respective physical, social, thinking and cyber spaces. To note, the $p_i(x_i, t)$, $s_i(x_i, t)$, $t_i(x_i, t)$, and $c_i(x_i, t)$ here are related to action $x_i$, and accumulated over a period $t$. The total entropy production of humans with the inside and the outside could be expressed as follows:

$$dS = d_i S + \sum_{i \in \{1,2,3,...,n\}} \int_{t_1}^{t_2} \bigl(p_i(x_i,t) + s_i(x_i,t) + t_i(x_i,t) + c_i(x_i,t)\bigr)dt \qquad (4)$$

If someone gets sick with Cyber-Syndrome, the value of $dS$ would tend to increase with higher randomness and disorders. At this state, the inside entropy of people may present relatively stable, and we assume it as a constant. Thus, the aim of Cyber-Syndrome recovery is to reduce the value of $d_e S$, in other words, to find most useful actions that minimize the value of entropy change with the outside:

$$Z = \min \sum_{i \in \{1,2,3,...,n\}} \int_{t_1}^{t_2} \bigl(p_i(x_i,t) + s_i(x_i,t) + t_i(x_i,t) + c_i(x_i,t)\bigr)dt \qquad (5)$$

where there are internal relationships between different spaces, for instance, the entropy change in physical space may have influences on that in social, thinking and cyber spaces. Hence, we define four constants of $\alpha$, $\beta$, $\gamma$, and $\epsilon$ as the impact factors of respective spaces. The hidden influence between different spaces could be expressed as equation 6-9.

$$p'_i(x_i, t) = p_i(x_i, t) + \beta * s_i(p_i(x_i, t), t) + \gamma * t_i(p_i(x_i, t), t) + \epsilon * c_i(p_i(x_i, t), t) \tag{6}$$

$$s'_i(x_i, t) = s_i(x_i, t) + \alpha * p_i(s_i(x_i, t), t) + \gamma * t_i(s_i(x_i, t), t) + \epsilon * c_i(s_i(x_i, t), t) \tag{7}$$

$$t'_i(x_i, t) = t_i(x_i, t) + \alpha * p_i(t_i(x_i, t), t) + \beta * s_i(t_i(x_i, t), t) + \epsilon * c_i(t_i(x_i, t), t) \tag{8}$$

$$c'_i(x_i, t) = c_i(x_i, t) + \alpha * p_i(c_i(x_i, t), t) + \beta * s_i(c_i(x_i, t), t) + \gamma * t_i(c_i(x_i, t), t) \tag{9}$$

where $\alpha + \beta + \gamma + \epsilon = 1$.

In addition, the entropy in each space is not infinitely high or low and needs to be considered in light of the actual situation. For example, people cannot participate in intensive physical exercise 24 hours a day for entropy reduction. There should exist a threshold $\theta$ of entropy change in respective space, taking the real limits into considerations:

$$|p'_i(x_i, t)|, |t'_i(x_i, t)|, |c'_i(x_i, t)|, |s'_i(x_i, t)| < \theta \tag{10}$$

During the recovery of Cyber-Syndrome, the final target is to reduce the total entropy production, so as to reach a balance of healthy state. According to equation 5, we need to find the most influential actions that could reduce the entropy to the utmost. Usually, the principle for Cyber-Syndrome recovery is to absorb lower entropy from the outside and release higher entropy to it. Here in Table II, we list some typical methods in respective spaces for Cyber-Syndrome recovery, which may provide guidance in practical applications.

TABLE II. The Potential Methods of Cyber-Syndrome Recovery in CPST Space.

| Spaces | Potential methods |
| --- | --- |
| Physical space | Keep adequate sleep and rest; Supplement nutrition; Stop eating junk food; Sunbathe; Take exercise; Speed up metabolism; Keep living environment clean and bright; Read more books and absorb useful information; Take medicine as prescribed by doctors... |
| Social space | Take part in various social activities; Establish healthy and harmonious relationships with surroundings; Abandon confused personal relationships... |
| Thinking space | Absorb advanced ideas and viewpoints; Get rid of outdated and backward concepts; Keep good moods; Seek and pursue the peace of mind; Ask others for help... |
| Cyberspace | Specify the time spent in cyberspace to avoid time fragmentation; Learn to identify and filter information online effectively; Take the advantages of the Internet and avoid any harmful information or temptation… |

## VI. CONCLUSIONS AND FUTURE WORKS

With the overwhelming development of the Internet and cyberspace, Cyber-Syndrome has become a new emerging disease which mainly stems from the excessive interactions with cyberspace. Based on our previous work, we continue to analyze the etiology of Cyber-Syndrome from the novel perspective of CPST space and Maslow's Hierarchy of Needs. In addition, cyber-enabled symptoms in physical, social, and thinking spaces have been depicted clearly. Moreover, we propose an entropy-based recovery mechanism of Cyber-Syndrome, which may provide guidance in practical applications. The unhealthy use of Internet makes more humans suffer from Cyber-Syndrome. It is of extreme significance to learn more about its etiology, symptoms, and recovery methods, so to achieve effective prevention and treatment.

Up to now, the research of Cyber-Syndrome is still in its infancy. Many existing works are related to the general introduction and dissemination of the concept, with little in-depth research on disease prevention and early detection. In the future, we aim to find ways of Cyber-Syndrome early detection via online social media, including text, image, video, and audio analysis etc. Moreover, combined with physiological signals and psychological features analysis, we will try to explore more intelligent ways for higher efficiency.